\documentclass[fleqn,10pt]{wlscirep}
\usepackage[utf8]{inputenc}
\usepackage[T1]{fontenc}
\usepackage{lineno}
\usepackage{array,multirow,booktabs,tabularx}
\usepackage{amsmath}
\usepackage{algorithm}
\usepackage{amssymb}   
\usepackage{pifont}

\usepackage[noend]{algpseudocode}
\algrenewcommand\algorithmiccomment[1]{\hfill\(\triangleright\) #1}

\title{MAD-LEO: A Maneuver-Annotated Orbital Dataset for LEO Satellites with Tiered Multi-Source Evidence}

\author[1,2,3,*]{Zhixin Guo}
\author[1,2,3,*]{Qi Shi}
\author[1,2,3,4,*,$\dag$]{Xiaofan Xu}
\author[1,2,3]{Linqiang Ge}
\author[1,2,3,$\dag$]{Hua Zhu}
\author[1,2,3]{Liyan Ben}
\author[4]{Bendian Nie}
\author[1,2,3]{Yuanrui Zhao}
\author[1,2,3]{Xiaohan Li}

\affil[1]{Shanghai Satellite Network Research Institute Co., LTD., Shanghai, 201210, China}
\affil[2]{State Key Laboratory of Satellite Network, Shanghai, 201210, China}
\affil[3]{Shanghai Key Laboratory of Satellite Network, Shanghai, 201210, China}
\affil[4]{China Satellite Network Innovation Co., LTD., Beijing, 100029, China}
\affil[*]{these authors contributed equally to this work}
\affil[$\dag$]{corresponding author(s): Xiaofan Xu (xiaofanxu@sina.com), Hua Zhu (zhuhua925@126.com)}

\begin{abstract}
With the rapid development of aerospace technology and the large-scale deployment of low Earth orbit (LEO) constellations, the risk of orbital collisions has increased, creating a growing demand for reliable observations of satellite maneuvers. However, public datasets containing real maneuver records remain scarce. We present MAD-LEO, a \textbf{M}aneuver-\textbf{A}nnotated orbital \textbf{D}ataset for \textbf{LEO} satellites. The mission-reported subset contains 1,134 maneuver events from eleven geodetic and altimetry satellites spanning 1992 to 2026, with labels taken directly from mission-published maneuver histories. Each event is checked against two-line element (TLE) data, precise orbit products, and satellite laser ranging (SLR) observations, with evidence tiers assigned according to data availability. The operational subset pairs operator-published ephemerides for 6,785 Starlink satellites with cataloged TLE records over a continuous 107-hour period. Technical validation across seven machine-readable experiment suites confirms the cross-source consistency of the labels and the evidence products.
\end{abstract}

\begin{document}

\flushbottom
\maketitle

\thispagestyle{empty}

\section*{Background \& Summary}

Low Earth orbit (LEO) is becoming increasingly crowded. Companies such as SpaceX, OneWeb, and Amazon are deploying large communication constellations. By the end of 2025, SpaceX had launched more than 9,000 Starlink satellites \cite{yamamoto2026tomography,figaro2026experimental}. Constellations at this scale requires frequent orbit maintenance and collision avoidance, and Starlink alone performs tens of thousands of avoidance maneuvers during each six-month reporting period \cite{parker2026data}.

The continued growth of these constellations has made the LEO environment increasingly dynamic. Satellite failures, re-entry, and environmental disturbances can further change the orbital population \cite{grile2025statistical,shirobokov2021survey,caldas2024machine}. At the same time, the increasing number of satellites creates challenges for astronomical observations and raises the risk of orbital collisions \cite{tao2022satellite}. Accurate and timely information on satellite maneuvers is therefore important for collision-risk assessment. A single unreported maneuver can invalidate the predicted trajectory used in conjunction screening.

Classical studies usually treat maneuver detection as an orbit-determination problem and rely on tracking observations \cite{vallado2001fundamentals}. The Extended Kalman Filter (EKF) \cite{smith1962application,julier1997new,einicke2012robust} has long been used for orbit estimation, while future trajectories can be obtained using analytical \cite{vallado2013improved,miura2009comparison} or numerical propagation methods \cite{urrutxua2016dromo,sharma1988long,aristoff2014orbit,bai2011modified,bradley2012new}. These approaches usually assume a known dynamical model and are often evaluated using simulated trajectories with predefined maneuvers.

Data-driven methods have been explored to reduce this dependence on assumed models. Two-line element set (TLE) data have been widely used as pseudo-observations to improve special-perturbations orbit propagation \cite{levit2011improved,bennett2012improving,sang2017analytical,san2017hybrid,peng2020machine,muldoon2009improved,peng2018exploring,peng2019comparative,peng2019gaussian,peng2021fusion}. Machine learning methods have also been used to correct the prediction errors of numerical propagation and simplified dynamical models. However, most of these studies focus on orbit prediction rather than maneuver detection. The few existing detection methods rely on particle-filter anomaly scoring applied to TLE data or on hybrids of unscented Kalman filtering and reachability analysis developed for sparse radar tracking \cite{rautalin2017latent,li2021improved,pihlajasalo2018improvement,san2018hybrid,curzi2022two,salleh2021adaptation,li2020machine}. Applying data-driven methods to maneuver detection requires labeled records of real maneuvers, which remain scarce in public orbital data — a limitation these detection studies explicitly note.

This problem becomes more important as the LEO population continues to grow while tracking capability remains limited. The most widely available public orbital data are TLEs distributed through Space-Track \cite{blasch2022space}. Their limited temporal resolution and lack of covariance information restrict their use for precise orbit tracking and maneuver analysis, particularly for frequently maneuvering satellites \cite{liu2024maneuver}. As a result, many studies rely on simulated data, which cannot fully represent real operational conditions \cite{tipaldi2022reinforcement}. Higher-accuracy orbit products are available for only a limited number of geodetic and altimetry missions. For these satellites, Global Navigation Satellite System (GNSS) and Doppler Orbitography and Radiopositioning Integrated by Satellite (DORIS) based precise orbit determination can achieve centimeter-level accuracy and can be independently checked using satellite laser ranging (SLR) \cite{tapley1994precision,selvan2023precise,degnan1993millimeter,arnold2019satellite}. For the broader satellite catalog, however, reliable maneuver labels remain scarce. Most available labels either lack independent evidence or are themselves the output of detection methods. They therefore cannot serve as ground truth for evaluation.

Operator-published ephemerides provide another source of orbital information. SpaceX publishes Starlink ephemerides through Space-Track with higher temporal resolution and better positional accuracy than TLE data, but these ephemerides are predicted trajectories rather than real-time observations and may therefore differ from the actual satellite motion \cite{liu2024maneuver}. The closest prior resource is the wide-scale monitoring of satellite dataset \cite{shorten2023wide}, which pairs long TLE histories for fifteen satellites with curated maneuver timestamps and has been used to evaluate maneuver-detection methods. However, its maneuver annotations are provided as event timestamps without event-level supporting evidence from independent orbital data sources. The operational subset covers a different setting, where validated maneuver records for individual satellites are not publicly available. It contains operator-published ephemerides for 6,785 Starlink satellites and the corresponding TLE records over a continuous 107-hour period in November 2024. These data allow maneuver-detection methods to be tested at scale under real operational conditions.

\section*{Methods}
\begin{figure*}[!t]
\centering
\includegraphics[scale=0.3]{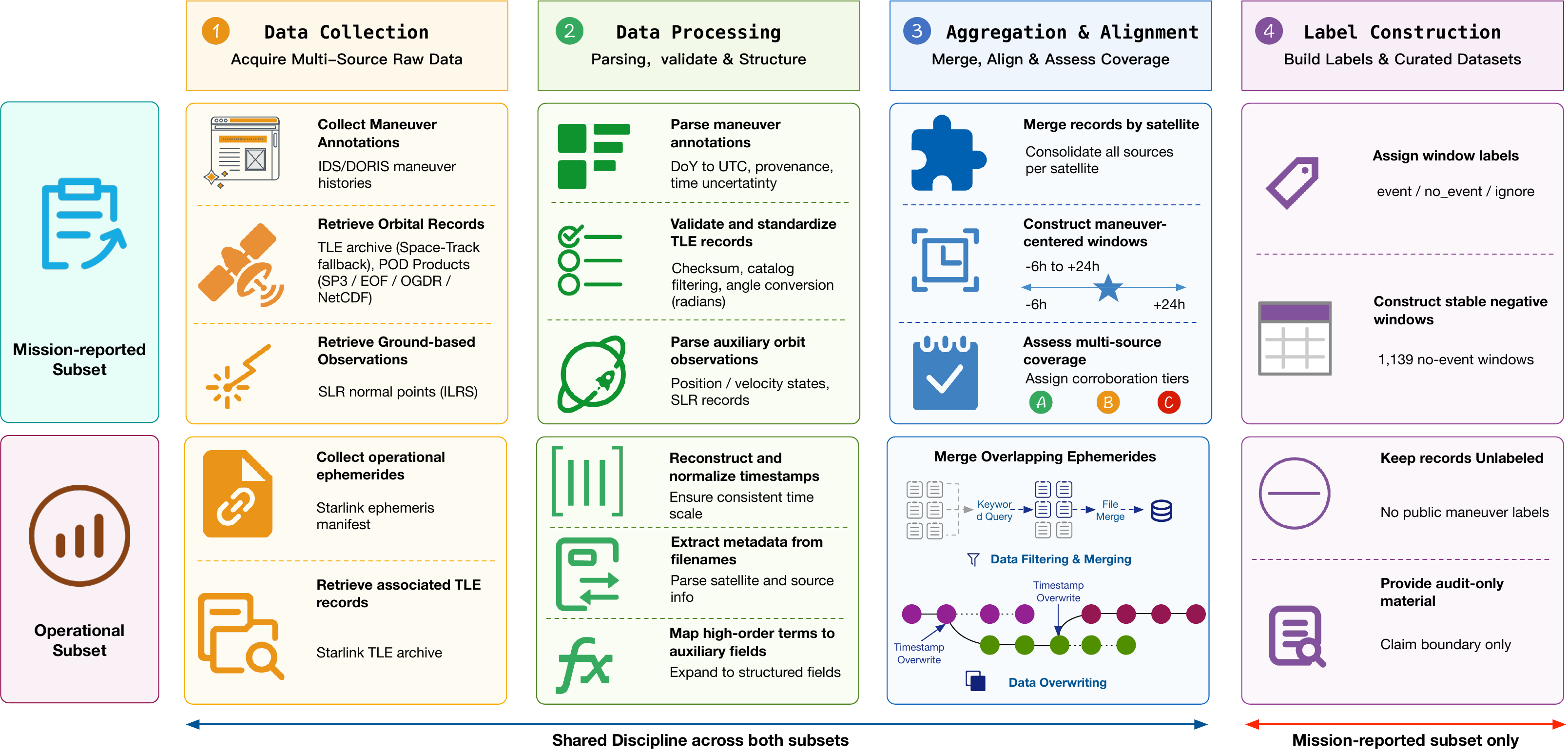}
\caption{Overview of the MAD-LEO dataset construction pipeline. MAD-LEO is constructed through four stages: data collection, data processing, aggregation and alignment, and label construction. The mission-reported subset integrates maneuver histories with TLE, precise-orbit, and ground-based ranging observations, whereas the operational subset combines operator-published ephemerides with associated TLE records. Both subsets follow a common framework for parsing, standardization, and satellite-level aggregation, with subset-specific procedures applied during alignment: mission-reported records are organized into maneuver-centered windows and assessed for multi-source coverage, while overlapping operational ephemerides are merged and checked against corresponding TLE records. Event, no-event, and ignore form the public label vocabulary of the mission-reported subset; the operational subset remains unlabeled because validated public maneuver annotations are unavailable.}
\label{dataconstruction}
\end{figure*}

As illustrated in Figure~\ref{dataconstruction}, MAD-LEO is constructed through a four-stage pipeline comprising data collection, data processing, aggregation and alignment, and label construction. During data collection, maneuver annotations and multi-source orbital observations are acquired for the mission-reported subset, while operator-published ephemerides and associated TLE records are collected for the operational subset. The collected data are then parsed, validated, and standardized before aggregation at the satellite level. Mission-reported records are further aligned to maneuver-centered analysis windows and assessed for multi-source coverage, whereas overlapping operational ephemerides are merged and checked against the corresponding TLE records. Event, no-event, and ignore form the public label vocabulary of the mission-reported subset; the operational subset remains unlabeled because no validated public maneuver annotations are available.

Table~\ref{tab:satellites} lists the satellites covered by the dataset. The mission-reported subset covers eleven geodetic and altimetry satellites. All eleven carry DORIS receivers and have mission-published maneuver histories available through the International DORIS Service (IDS). They are also tracked by the International Laser Ranging Service (ILRS). The operational subset covers the Starlink constellation.

\begin{table}[ht]
\centering
\caption{Satellites in MAD-LEO and their available data sources. ``\checkmark'' and ``\ding{55}'' indicate whether the corresponding source is included in the dataset. The precise-orbit column gives the product used for each mission; all listed products are precise-orbit grade except the Jason-3 OGDR, which is a near-real-time operational product, and the Sentinel-6A entry (RINEX), which denotes daily GNSS tracking files used only to establish orbit-data coverage rather than a processed orbit product. Ephemeris refers to operator-published predicted ephemerides. First and last maneuver indicate the dates of the first and last mission-reported events. The Starlink row represents the operational subset and contains no validated maneuver labels.}
\label{tab:satellites}
\setlength{\tabcolsep}{3pt}
\footnotesize
\begin{tabular}{lccccccl}
\hline
Satellite & TLE & Precise orbit product & SLR & Ephemeris & First maneuver & Last maneuver \\
\hline
TOPEX/Poseidon & \checkmark & SP3 & \checkmark & \ding{55} & 1992-08-17 & 2004-11-17 \\
Jason-1 & \checkmark & SP3 & \checkmark & \ding{55} & 2001-12-11 & 2013-06-13 \\
Jason-2 & \checkmark & SP3 & \checkmark & \ding{55} & 2008-06-23 & 2019-10-04 \\
CryoSat-2 & \checkmark & SP3 & \checkmark & \ding{55} & 2010-04-15 & 2026-07-01 \\
HY-2A & \checkmark & SP3 & \checkmark & \ding{55} & 2011-09-28 & 2020-06-09 \\
SARAL/AltiKa & \checkmark & SP3 & \checkmark & \ding{55} & 2013-02-27 & 2026-02-08 \\
Jason-3 & \checkmark & OGDR & \checkmark & \ding{55} & 2016-01-19 & 2026-07-05 \\
Sentinel-3A & \checkmark & AUX\_POEORB & \checkmark & \ding{55} & 2016-02-22 & 2026-06-18 \\
Sentinel-3B & \checkmark & AUX\_POEORB & \checkmark & \ding{55} & 2018-04-30 & 2026-06-11 \\
Sentinel-6A & \checkmark & RINEX & \checkmark & \ding{55} & 2020-11-23 & 2026-05-20 \\
SWOT & \checkmark & NetCDF & \checkmark & \ding{55} & 2023-01-11 & 2026-06-04 \\ \hline
Starlink satellites & \checkmark & \ding{55} & \ding{55} & \checkmark & -- & -- \\
\hline
\end{tabular}
\end{table}

\textbf{Data Collection}. Four types of source data were collected for the eleven reference satellites in the mission-reported subset: maneuver annotations, TLE records, precise-orbit products, and SLR observations. Maneuver annotations were obtained from maneuver history files published by the International DORIS Service (IDS) \cite{tavernier2005international}. These files document orbit adjustments for DORIS-equipped geodetic and altimetry missions and report the first-impulse time together with the operation start and end times when available.

TLE records were obtained primarily from non-interactive public archives (CelesTrak), with the authenticated Space-Track general perturbations history interface used when archival coverage was unavailable \cite{blasch2022space,kelso2017challenges}. For each satellite, TLE records were collected over a window extending up to 30 days before the first reported maneuver and up to 30 days after the last, clipped to the satellite's primary coverage period.

Precise-orbit data were obtained from the Crustal Dynamics Data Information System (CDDIS) of the National Aeronautics and Space Administration (NASA), the Physical Oceanography Distributed Active Archive Center (PO.DAAC), and the Copernicus Data Space Ecosystem (CDSE) \cite{noll2010crustal,d2026copernicus,moroni2016managing}. Depending on mission availability, the products were provided in Standard Product 3 (SP3), Earth Observation Format (EOF), Operational Geophysical Data Record (OGDR), or Network Common Data Form (NetCDF) formats \cite{weiss2017orbit,yang2006review,biancamaria2018validation,rew1990netcdf}. The product selected for each mission is listed in Table~\ref{tab:satellites}. Sentinel-6A was treated separately because its publicly available orbit data consist of daily GNSS tracking files in Receiver Independent Exchange Format (RINEX) rather than a processed orbit product \cite{montenbruck2021sentinel}. These files were used to establish the available orbit-data coverage.

SLR observations were obtained from the ILRS archives hosted by the CDDIS \cite{noll2010crustal} and the European Laser Consortium (EUROLAS) Data Center \cite{schwatke2016eurolas}. The released SLR records contain both normal-point and full-rate observations, with normal-point precision retained where available.

The operational subset was constructed from Starlink operator-published predicted ephemerides available through Space-Track \cite{blasch2022space,liu2024maneuver}. Ephemerides were collected continuously from 06:00 Coordinated Universal Time (UTC) on 26 November to 17:00 UTC on 30 November 2024, defining a 107-h observation interval. TLE records covering the same period were obtained from public archives, with Space-Track used where necessary to supplement archival coverage. Retrieval from Space-Track was performed in accordance with the platform terms of use.

All source material for the mission-reported subset was retrieved up to 13 August 2026. The IDS maneuver histories are updated on a rolling basis. The released annotations therefore correspond to this snapshot, and the retrieval date should be quoted when comparing label sets across studies. Two-line elements are factual catalog data distributed by the United States Space Command through public archives and the Space-Track service. The released TLE tables redistribute these records with attribution and modify them only through parsing and unit normalization. The Starlink ephemerides are prediction files published openly by the operator for conjunction screening, and they are redistributed here for the same purpose. Users who access the underlying services themselves must comply with the Space-Track user agreement. Raw provider files subject to provider-specific redistribution terms (precise-orbit products, SLR observations, and RINEX tracking files) are not redistributed. The released evidence snapshots contain only the parsed observation records needed to reproduce the analyses.

\textbf{Data Processing}. Data from the individual sources were parsed and converted into standardized records. For maneuver annotations, the first-impulse time was used as the representative maneuver epoch when available; otherwise, the reported operation end time was used. The reported operation interval was retained where available.

TLE records from both subsets were validated using the checksum procedure recommended by CelesTrak \cite{kelso2017challenges} and cross-checked against the corresponding satellite catalog, with records carrying inconsistent North American Aerospace Defense Command (NORAD) identifiers removed. Precise-orbit products were parsed into position and velocity state vectors, retaining the accuracy information provided with the original products. SLR observations were represented as range records in meters. Normal-point and full-rate observations were stored as separate record types, and precision and residual information were retained where available. Starlink ephemeris timestamps were reconstructed from the reported start epoch and sampling interval.

Time references were standardized to UTC \cite{panfilo2019coordinated}. The DORIS SP3 products from Segment Sol multimissions d'ALTimétrie, d'Orbitographie et de localisation précise (SSALTO) \cite{pujol2013ssalto} and the Groupe de Recherche de Géodésie Spatiale (GRG) analysis center\cite{forste2008geoforschungszentrum} use International Atomic Time (TAI), while Goddard Space Flight Center (GSFC) products \cite{center1991goddard} use Global Positioning System (GPS) time. These timestamps were converted to UTC using the corresponding leap-second records. Maneuver times reported as year and day of year were converted directly to UTC with leap years taken into account.

For cross-source comparison in the mission-reported subset, TLE-derived states, originally represented in the True Equator Mean Equinox (TEME) frame \cite{seago2000coordinate}, were propagated using SGP4 and transformed to the International Terrestrial Reference Frame (ITRF) \cite{dong2003origin}. Positions and ranges were expressed in meters, velocities in meters per second, angular elements in radians, and mean motion in radians per minute. Source provenance was retained with the processed records to preserve traceability. Orbital calculations use the WGS-84 gravitational parameter $\mu = 3.986004418\times10^{14}$~m$^3$/s$^2$; reported altitudes follow a single release-wide convention that subtracts the spherical mean Earth radius 6378.137~km from the orbital radius, and SGP4 propagation uses its internal WGS-72 constants.

\textbf{Aggregation and Alignment}. Processed records were first grouped by satellite. Records from the different sources were then aligned in time. For the mission-reported subset, TLE, precise-orbit, and SLR records were associated with the corresponding satellite over its available mission coverage period. The operational subset required an additional deduplication step. Starlink ephemerides consist of overlapping rolling predictions, and the published files carry no publication timestamp that identifies the most recent prediction. Duplicate epochs were therefore resolved deterministically by filename order. Finally, the temporal overlap between each ephemeris series and the corresponding TLE records was checked before cross-source comparison.

Event alignment was applied only to the mission-reported subset, as the operational subset contains no reported maneuvers. Each reported maneuver was assigned a 30-h analysis window, extending from 6~h before to 24~h after the representative maneuver epoch. The window is asymmetric so that the post-maneuver portion can capture changes that may enter catalog records only after the reported maneuver time. The original mission-reported operation interval was retained independently of this window.

Data availability within each window was then assessed separately for the three sources. TLE coverage required at least one catalog epoch at or before the window start and at least one epoch at or after its end. Precise-orbit coverage required temporal overlap between the orbit product and the window. For Sentinel-6A, where no processed orbit product exists, the daily RINEX tracking files were used to establish orbit-data coverage instead. SLR coverage was evaluated over an extended interval that adds one day on either side of the window, to account for the sparse, pass-based sampling of laser-ranging observations.

Based on this assessment, each maneuver window was assigned a confidence tier that reflects the extent of independent multi-source evidence supporting the reported maneuver. Tier~A denotes windows with TLE, precise-orbit, and SLR data all available. Tier~B denotes windows with TLE and precise-orbit data available but insufficient or missing SLR observations. Tier~C denotes windows in which one or more principal sources are unavailable. For Sentinel-6A, the availability of the daily RINEX tracking files takes the orbit-coverage role in this assignment. Its windows do not enter any of the orbit-derived response analyses reported in Technical Validation, as no processed orbit states exist for this target. The confidence tier therefore characterizes the multi-source observational support of an event, not the reliability of the mission-reported maneuver label itself.

\begin{table}[!htbp]
\centering
\footnotesize
\renewcommand{\arraystretch}{1.02}
\emergencystretch=1.5em
\begin{tabularx}{\textwidth}{@{}p{0.32\textwidth}ccX@{}}
\hline
\textbf{File name on the deposit} & \textbf{Files} & \textbf{Format} & \textbf{Contents} \tabularnewline
\hline
\texttt{docs\_\_metadata.md} & 1 & Markdown & Column dictionaries, label protocol, and data-quality notes \tabularnewline
\multicolumn{4}{@{}l@{}}{\texttt{mission\_reported\_\_annotations\_\_}} \tabularnewline
\texttt{maneuver\_annotations.csv} & 1 & CSV & Mission-reported maneuver records with verbatim source lines (1,134 rows) \tabularnewline
\texttt{event\_windows.csv} & 1 & CSV & Event windows with per-source coverage and confidence tiers (1,134 rows) \tabularnewline
\texttt{stable\_windows.csv} & 1 & CSV & No-event control windows (1,139 rows) \tabularnewline
\texttt{stable\_windows\_matched.csv} & 1 & CSV & Coverage-matched no-event subset (274 rows) \tabularnewline
\multicolumn{4}{@{}l@{}}{\texttt{mission\_reported\_\_evidence\_\_tle\_\_}} \tabularnewline
\texttt{<sat\_id>.parquet} & 11 & Parquet & Full-mission TLE element histories, one file per satellite (3,495--15,355 rows each) \tabularnewline
\multicolumn{4}{@{}l@{}}{\texttt{mission\_reported\_\_evidence\_\_orbit\_\_}} \tabularnewline
\texttt{<sat\_id>.parquet} & 10 & Parquet & Precise-orbit state series, one file per satellite except Sentinel-6A \tabularnewline
\multicolumn{4}{@{}l@{}}{\texttt{mission\_reported\_\_evidence\_\_slr\_\_}} \tabularnewline
\texttt{<sat\_id>.parquet} & 11 & Parquet & SLR observations, $\sim$2.1M normal points and $\sim$24M full-rate records in total \tabularnewline
\multicolumn{4}{@{}l@{}}{\texttt{operational\_\_starlink\_\_}} \tabularnewline
\texttt{ephemeris\_state.parquet} & 1 & Parquet & Starlink predicted states (43,361,358 rows) \tabularnewline
\texttt{ephemeris\_state\_sample25.parquet} & 1 & Parquet & 25-satellite sample for quick inspection of the data structure \tabularnewline
\texttt{tle\_elements.parquet} & 1 & Parquet & Starlink TLE records over the same interval (51,475 epochs) \tabularnewline
\hline
\end{tabularx}
\caption{Files of the release, named as deposited. A row in the first column set in full width gives the file-name prefix shared by the files listed below it, and \texttt{<sat\_id>} stands for the satellite identifiers (e.g., \texttt{sentinel-3a.parquet}). The double underscores encode the logical directory structure explained in the text.}
\label{tab:dr_inventory}
\end{table}

The aligned windows were also used to derive source-specific response measures for cross-source comparison. For TLE data, the response was defined as the change in semi-major axis computed from the mean motion of the nearest catalog epochs before and after the window using Kepler's third law \cite{russell1964kepler}. Angular elements were unwrapped before differences were calculated. For precise-orbit data, the response was defined as the difference in period-averaged semi-major axis between the 12-h intervals immediately before and after the window. A one-nodal-period rolling average was applied to the osculating semi-major-axis series to suppress short-period variations before comparison with the TLE-derived change. For SLR data, normal-point precision statistics were summarized within the window and within the 12-h intervals immediately before and after it. An SLR response was calculated only when both surrounding intervals contained at least two valid observations. Windows with insufficient observations were retained but received no SLR response.

\textbf{Label Construction}. Mission-reported maneuvers were used directly as positive events. Orbital changes derived from TLE, precise-orbit, or SLR data were not used to define maneuver labels. An ignore label is reserved for records that cannot be assigned a valid UTC epoch or that carry an unresolved conflict with their source record. Insufficient evidence coverage is never grounds for ignoring a mission-reported event. Such events are retained, and the missing evidence is expressed through the per-source coverage status and the confidence tier. In this release, every parsed mission record is released as an event, and no window carries the ignore label. The value remains part of the public label vocabulary so that future audit extensions can adopt it without schema changes.

No-event windows were then generated for each satellite on a non-overlapping 30-h grid beginning at the first available TLE epoch. Candidate windows overlapping a maneuver window extended by 24~h on either side were excluded. Within each calendar year, the number of no-event windows was matched to the number of reported maneuvers, with at least one window whenever eligible candidates were available. Windows were selected at approximately even intervals across the available candidates. Event and no-event windows were assessed with the same source-coverage rules and assigned confidence tiers under the same criteria.

A matched subset of no-event windows was further constructed so that the two window types would be comparable in data support. For each satellite, the fraction of maneuver events covered by all three sources was computed from the coverage assessment. The no-event windows of that satellite were divided into a Tier A pool and a pool of the remaining windows. The subset was then deterministically resampled so that its Tier A share reproduced the event fraction to the nearest integer, subject to the available pool sizes. Satellites were skipped when they had no Tier A pool of no-event windows or no fully covered maneuver event. This procedure yielded 274 matched no-event windows.

Maneuver candidates identified from orbital data during internal analysis were likewise excluded from the label set. The operational Starlink subset contains no independently validated maneuver annotations and therefore has no maneuver ground-truth labels.

\section*{Data Records}
MAD-LEO is publicly available through Figshare \cite{madleo_figshare} under the Creative Commons Attribution 4.0 International license (CC BY 4.0). The release contains 40 files with a total size of approximately 4.3 GB. Label tables are stored as comma-separated value (CSV) files, and the orbital records as Zstandard-compressed Parquet files. Table~\ref{tab:dr_inventory} lists every deposited file with its format and contents, and files are referred to by their contents throughout this section. All timestamps are ISO-8601 UTC, and complete column dictionaries are provided in the metadata document on the deposit.

\begin{table}[!htbp]
\centering
\footnotesize
\renewcommand{\arraystretch}{1.02}
\emergencystretch=1.5em
\begin{tabularx}{\textwidth}{@{}p{0.15\textwidth}p{0.33\textwidth}X@{}}
\hline
\textbf{Category} & \textbf{Field} & \textbf{Description} \tabularnewline
\hline
Identification &
\texttt{annotation\_id}, \texttt{sat\_id} &
Unique identifier of the event or window, and satellite identifier used throughout the release. \tabularnewline
Timing &
\texttt{event\_time\_utc} &
Timestamp used to anchor the analysis window. The first-impulse time is used when available and the reported operation end time otherwise. \tabularnewline
&
\texttt{window\_start\_utc}, \texttt{window\_end\_utc} &
Start and end times of the 30-h analysis window. \tabularnewline
&
\texttt{time\_uncertainty\_seconds} &
Timing uncertainty recorded for the annotation. \tabularnewline
Label and coverage &
\texttt{event\_label} &
Window label, given as \texttt{event}, \texttt{no\_event}, or \texttt{ignore}. \tabularnewline
&
\texttt{confidence\_tier} &
Evidence-coverage class A, B, or C. \tabularnewline
&
\texttt{tle\_status}, \texttt{slr\_status}, \texttt{orbit\_status} &
Coverage status of each data source within the analysis window. \tabularnewline
&
\texttt{aligned}, \texttt{missing\_sources} &
Whether all three sources satisfy the alignment criteria, and the missing sources when full coverage is unavailable. \tabularnewline
Mission record &
\texttt{raw\_record} &
Original source record from the mission history. \tabularnewline
&
\texttt{reported\_operation\_start\_utc}, \texttt{reported\_operation\_end\_utc} &
Start and end times of the maneuver operation reported by the mission. \tabularnewline
&
\texttt{event\_type}, \texttt{event\_time\_role} &
Event class and the source used to define \texttt{event\_time\_utc}. \tabularnewline
&
\texttt{impulse\_count}, \texttt{mission\_record\_code} &
Number of reported maneuver impulses and the mission or file identifier recorded in the source data. \tabularnewline
Provenance and quality &
\texttt{annotation\_status}, \texttt{truth\_status} &
Status fields indicating that the label is mission-reported rather than generated by a detection algorithm. \tabularnewline
&
\texttt{source}, \texttt{source\_provider}, \texttt{source\_url}, \texttt{reference} &
Source document, provider, URL, and citation information. \tabularnewline
&
\texttt{batch}, \texttt{scope} &
Planning batch and window scope recorded for the entry. \tabularnewline
&
\texttt{annotation\_label\_source}, \texttt{quality\_flags}, \texttt{notes} &
Label source, machine-readable quality flags, and processing notes. \tabularnewline
\hline
\end{tabularx}
\caption{Fields of the annotation and window tables. Complete field definitions are provided in the metadata document on the deposit.}
\label{tab:dr_annotations}
\end{table}

The deposit stores its files as a flat list, without directory hierarchy. The logical structure of the release is therefore encoded in the file names, with path separators written as double underscores: the deposited file \textit{mission\_reported\_\_evidence\_\_tle\_\_sentinel-3a.parquet} corresponds to the logical path \textit{mission\_reported/evidence/tle/sentinel-3a.parquet}. The metadata document follows the same convention.

\subsection*{Mission-reported subset}

Four CSV tables form the annotation layer (Table~\ref{tab:dr_inventory}). The annotation table holds the 1,134 mission-reported maneuvers, and each record retains the verbatim source line so that every label can be traced to the IDS/DORIS maneuver history it came from. The event-window table links each maneuver to its 30-h analysis window and records the per-source coverage status and the confidence tier. The stable-window table provides the 1,139 no-event control windows constructed in Methods, and the matched subset holds the 274 windows recommended for balanced evaluation. Fourteen stable windows carry a \texttt{suspect\_unreported\_maneuver} flag because their TLE-derived response exceeds 20~m, a plausible sign of an unreported orbit change. These windows are retained, and Technical Validation examines them individually.

Most annotations anchor the window at the reported first-impulse time (1,089 of 1,134). The remaining 45 use the reported operation end time, the only epoch those records carry. No released window carries the \texttt{ignore} label. The value remains part of the public vocabulary for records that cannot be assigned a valid UTC epoch or that conflict with their source. Applying the tier criteria of Methods classifies 754 events as tier A, 194 as tier B, and 186 as tier C. The no-event windows split into 204 tier A, 83 tier B, and 852 tier C. A tier records how much independent orbital evidence supports a window and never modifies the mission-reported label. Table~\ref{tab:dr_annotations} summarizes the fields of the annotation and window tables.

The orbital evidence is stored as one Parquet file per satellite and per source: eleven TLE files, ten precise-orbit files, and eleven SLR files (Table~\ref{tab:dr_inventory}). The TLE files cover all eleven reference satellites over their full mission spans and carry the catalog elements in radians. The precise-orbit files cover ten satellites in the ITRF, with positions in meters and Earth-fixed velocities in meters per second. Their sources are the DORIS SP3, Sentinel-3 AUX\_POEORB, Jason-3 operational GPS, and Surface Water and Ocean Topography (SWOT) \cite{fu2024surface} products listed in Table~\ref{tab:satellites}. Sentinel-6A has no orbit file: its orbit coverage was established from daily GNSS tracking files, which are not redistributed (Methods). The SLR files cover all eleven satellites and contain approximately 2.1 million normal points and 24 million full-rate observations from the ILRS archives, including legacy records in formats predating the Consolidated Ranging Data (CRD) standard \cite{pearlman2002international} for the earlier missions. The \texttt{record\_type} field distinguishes the two observation types.

For normal-point records, \texttt{sigma\_m} is the precision of the normal point, that is, the spread of the underlying range measurements. It is not an orbit residual. Some source-side placeholder values are kept verbatim: about 9,600 normal points carry a negative placeholder sigma, and a small tail exceeds 1~m. Twenty rows with implausible ranges or cross-target contamination are flagged in the \texttt{qc\_status} column rather than removed. The counts and the recommended filters are documented in the metadata document. Table~\ref{tab:dr_evidence} summarizes the fields of the three evidence types.

\subsection*{Operational subset}

The operational subset is a continuous record of constellation-scale Starlink operations rather than an event-centered one. The main ephemeris file contains 43,361,358 predicted states for 6,785 distinct NORAD catalog objects, derived from 6,761 operator-published ephemeris files. Twenty-four operator names are shared by two catalog objects, which accounts for the difference between the two counts. The records span the 107-h interval from 06:00 UTC on 26 November to 17:00 UTC on 30 November 2024 at a nominal 60-s cadence. The subset also includes a 25-satellite sample for quick inspection of the data structure and the TLE records covering the same interval (51,475 epochs).

Each state carries the NORAD catalog number, the operator catalog name, the UTC epoch, and position and velocity components in the MEME frame (Table~\ref{tab:dr_starlink}). The published predictions of three actively deorbiting objects continue below the Earth's radius, and the 5,861 affected rows (0.014\%) are flagged \texttt{below\_surface} in \texttt{quality\_flag}. They are retained verbatim, and altitude statistics should filter on this field.

\begin{table}[!htbp]
\centering
\footnotesize
\renewcommand{\arraystretch}{1.02}
\emergencystretch=1.5em
\begin{tabularx}{\textwidth}{@{}p{0.15\textwidth}p{0.33\textwidth}X@{}}
\hline
\textbf{Data type} & \textbf{Field} & \textbf{Description} \tabularnewline
\hline
TLE &
\texttt{sat\_id}, \texttt{epoch} &
Satellite identifier and catalog epoch in UTC. \tabularnewline
&
\texttt{mean\_motion\_rad\_per\_min} &
Mean motion in radians per minute. \tabularnewline
&
\texttt{inclination\_rad}, \texttt{eccentricity} &
Orbital inclination in radians and orbital eccentricity. \tabularnewline
&
\texttt{bstar\_per\_earth\_radius} &
Drag term in inverse Earth radii. \tabularnewline
Precise orbit &
\texttt{sat\_id}, \texttt{epoch} &
Satellite identifier and state epoch in UTC. \tabularnewline
&
\texttt{x\_m}, \texttt{y\_m}, \texttt{z\_m} &
Position components in the International Terrestrial Reference Frame, in meters. \tabularnewline
&
\texttt{vx\_mps}, \texttt{vy\_mps}, \texttt{vz\_mps} &
Velocity components in meters per second (Earth-fixed frame; add $\omega \times r$ before inertial-frame use). \tabularnewline
&
\texttt{sigma\_x\_m}, \texttt{sigma\_y\_m}, \texttt{sigma\_z\_m} &
Position uncertainty fields retained from the source product schema. All values are null in this release. \tabularnewline
&
\texttt{source\_product} &
Source file from which the state record was obtained. \tabularnewline
&
\texttt{clock}, \texttt{clock\_rate} &
Satellite clock correction and clock rate reported by the DORIS SP3 and Jason-3 OGDR products. \tabularnewline
&
\texttt{quality} &
Orbit quality flag reported by the Sentinel-3 AUX\_POEORB products. \tabularnewline
&
\texttt{orbit\_qual} &
Orbit quality flag reported by the SWOT product. \tabularnewline
SLR &
\texttt{record\_type} &
Observation type, either normal point or full rate. \tabularnewline
&
\texttt{epoch} &
Observation epoch in UTC. \tabularnewline
&
\texttt{time\_of\_flight\_s} &
Two-way time of flight in seconds. \tabularnewline
&
\texttt{range\_m} &
One-way range to the satellite in meters. \tabularnewline
&
\texttt{sigma\_m} &
Precision of the normal point in meters where applicable. \tabularnewline
&
\texttt{num\_returns} &
Number of individual measurements included in the normal point where applicable. \tabularnewline
&
\texttt{window\_length} &
Duration of the normal-point integration interval in seconds where applicable. \tabularnewline
&
\texttt{station\_id}, \texttt{target\_id} &
Station and ILRS satellite identifiers. \tabularnewline
&
\texttt{target\_name} &
Target name recorded in the observation file. \tabularnewline
&
\texttt{source\_zero\_fields} &
Marker for source-unfilled fields. The provider left \texttt{sigma} and/or \texttt{num\_returns} at zero, which is a structural zero rather than a measurement. \tabularnewline
&
\texttt{qc\_status} &
Row-level physical quality-control status (\texttt{ok}, \texttt{range\_implausible}, \texttt{cross\_target}, or \texttt{qc\_rejected}). Flagged rows are retained verbatim, and statistics should filter on this field. \tabularnewline
\hline
\end{tabularx}
\caption{Fields of the TLE, precise-orbit, and SLR evidence records.}
\label{tab:dr_evidence}
\end{table}

\begin{table}[!htbp]
\centering
\footnotesize
\renewcommand{\arraystretch}{1.02}
\emergencystretch=1.5em
\begin{tabularx}{\textwidth}{@{}p{0.30\textwidth}X@{}}
\hline
\textbf{Field} & \textbf{Description} \tabularnewline
\hline
\texttt{sat\_id} &
NORAD catalog number. \tabularnewline
\texttt{satellite\_name} &
Satellite name in the operator catalog. \tabularnewline
\texttt{epoch} &
State epoch in UTC at 60 s intervals. \tabularnewline
\texttt{x\_m}, \texttt{y\_m}, \texttt{z\_m} &
Position components in the MEME reference frame, in meters. \tabularnewline
\texttt{vx\_mps}, \texttt{vy\_mps}, \texttt{vz\_mps} &
Velocity components in meters per second in the MEME reference frame. \tabularnewline
\texttt{quality\_flag} &
Row-level quality marker. \texttt{below\_surface} marks the 5,861 states (0.014\%, from three actively deorbiting objects) whose operator-published predictions continue below the Earth's radius. Such rows are retained verbatim and flagged, and altitude statistics should filter on this field. \tabularnewline
\hline
\end{tabularx}
\caption{Fields of the Starlink ephemeris records.}
\label{tab:dr_starlink}
\end{table}

\section*{Technical Validation}

The validation of MAD-LEO comprises seven groups of experiments: annotation traceability, cross-source response consistency, a satellite-laser-ranging audit, the operational-subset boundary, harmonized state estimation, tier stratification, and evidence-distribution completeness. Each group tests one way in which the release could fail: labels that do not parse or do not match independent records, a response metric that does not measure maneuver magnitude, corrupted laser-ranging observations, misread operational predictions, unstable state estimates, tier-dependent label quality, and biased evidence sampling. Every experiment regenerates from machine-readable artifacts distributed with the code repository, and Table~\ref{tab:denominators} records the sample size and binding constraint of each analysis, so every denominator quoted below can be traced to a released table.

\begin{table}[htbp]
\centering
\footnotesize
\setlength{\tabcolsep}{3pt}
\renewcommand{\arraystretch}{1.02}
\begin{tabular}{p{0.30\textwidth}p{0.18\textwidth}p{0.44\textwidth}}
\hline
\textbf{Analysis} & \textbf{Sample} & \textbf{Binding constraint} \tabularnewline
\hline
TLE response (all events) & 1,134 & Computable for every released event window. \tabularnewline
Dual TLE--orbit comparison & 556 & Orbit state samples required in both 12-h bracketing bands (at least two samples per band, i.e. at least one period-averaged value each); Sentinel-6A excluded (tracking coverage only). \tabularnewline
SLR precision-shift audit & 542 & At least two normal points on each side of the window. \tabularnewline
SLR geometric O-C audit & 296 & At least three observations on each side of the window. \tabularnewline
Sigma-model fit and evaluation & 726 (719) & Event-epoch state estimates (266 fit, 195 contemporary, 265 historical); 719 inside the 48-h validity domain, 7 beyond. \tabularnewline
Tier-stratified checks & 558/168, 418/138 (A/B); 754/194/186 (A/B/C) & Orbit-based metrics (three-sigma coverage, sign agreement) computable for tiers A and B only, because tier C lacks precise-orbit evidence; the TLE response is computed for every tier. \tabularnewline
Kozai convention check & 1,134 & Recomputed for every event window. \tabularnewline
Stable-window specificity & 1,133 of 1,139 & Six windows at the first TLE grid point of a satellite lack bracketing catalog epochs. \tabularnewline
Matched-subset specificity & 273 of 274 & Same constraint within the coverage-matched subset. \tabularnewline
Window-sensitivity check & 1,134 & Every response recomputed under a symmetric $\pm$24-h window definition. \tabularnewline
Synthetic attenuation demo & 20 & Injection grid (5 magnitudes $\times$ 4 bracket spacings) on a quiet Sentinel-3A precise-orbit arc. \tabularnewline
External benchmark match & 687 of 742 & One-day tolerance, benchmark events recovered (709 of 742 at two days, 80 at a half day); 55 unmatched in-span, 392 mission-reported events outside benchmark coverage. \tabularnewline
\hline
\end{tabular}
\caption{Sample sizes of the technical-validation analyses. Each row records the number of windows entering the reported statistic and the constraint that determines it, so that every denominator quoted in this section can be traced to a released table.}
\label{tab:denominators}
\end{table}

\begin{figure*}[!t]
\centering
\includegraphics[width=\textwidth]{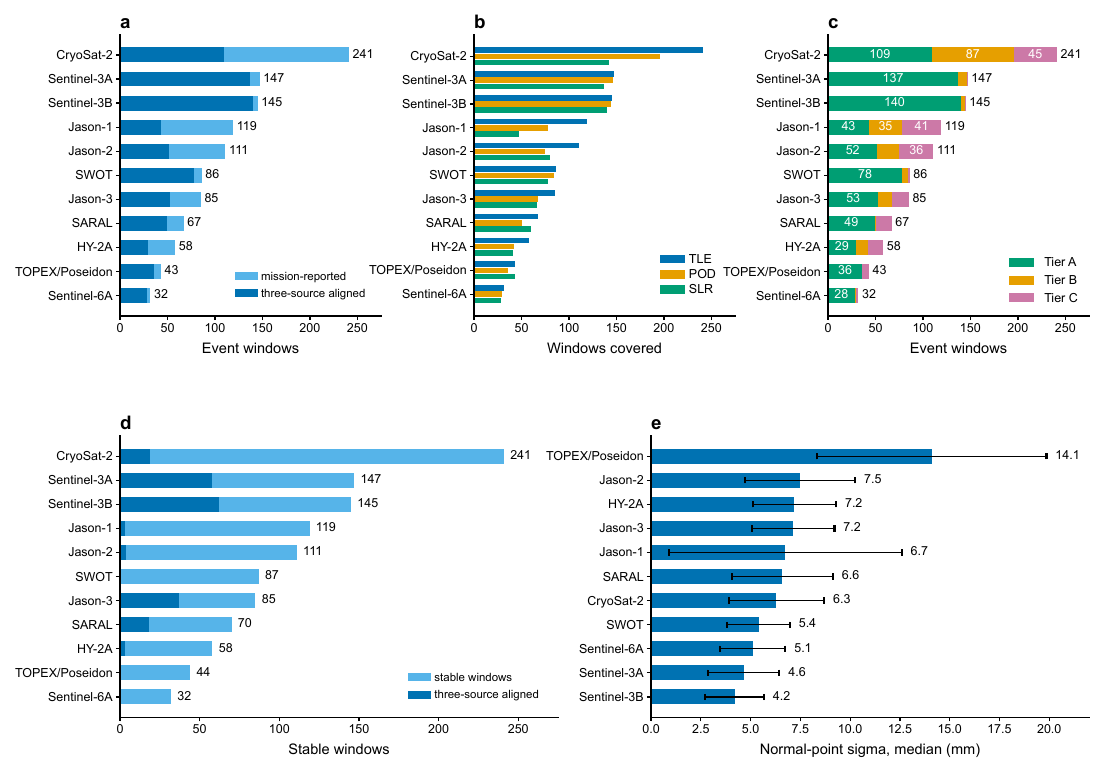}
\caption{Composition and coverage of the mission-reported subset. (a) Mission-reported event windows and the all-source-aligned subset per target. (b) Per-target coverage counts for the three evidence sources. (c) Confidence-tier composition of the 1,134 released events (A = 754, B = 194, C = 186). (d) Stable no-event control windows and their all-source-aligned subset per target. (e) Per-target SLR normal-point precision (median with inter-quartile range).}
\label{fig:dataset}
\end{figure*}

Figure~\ref{fig:dataset} summarizes the composition of the mission-reported subset. TLE and precise-orbit coverage are nearly complete for all eleven missions (panels a and b). SLR observes only during station passes and is the binding constraint on three-source coverage, with availability tracking the mission era. The confidence tiers record this heterogeneity as a per-window attribute: of the 1,134 released events, 754 carry full three-source corroboration (tier A), 194 lack SLR coverage (tier B), and 186 lack precise-orbit coverage (tier C) (panel c). The 1,139 stable control windows follow the same structure (204 tier A, 83 tier B, 852 tier C; panel d), and the release provides a coverage-matched subset of 274 windows for balanced evaluation. Panel (e) characterizes the laser-ranging instrument used in the audit below: per-target normal-point precisions of 4--14~mm, three orders of magnitude below the tens-of-meter responses under test.

\begin{figure*}[!t]
\centering
\includegraphics[width=\textwidth]{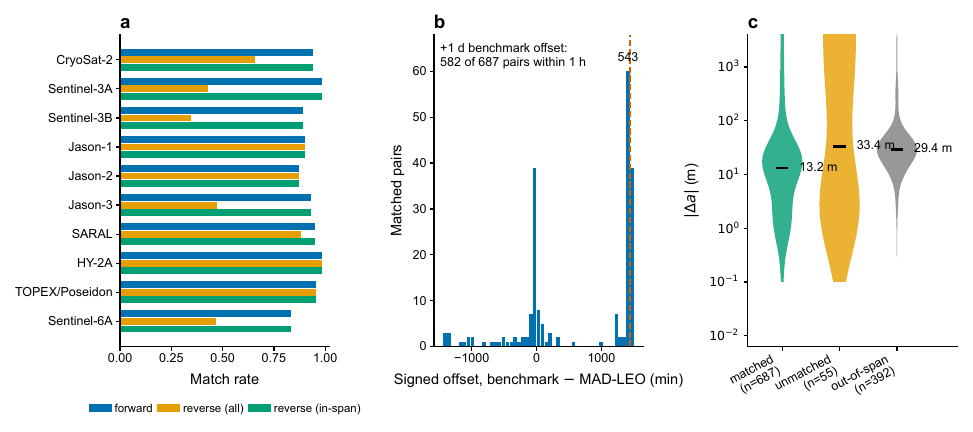}
\caption{Cross-check of the mission-reported annotations against the independently published maneuver benchmark of Shorten et al. (a) Match rates per target in both directions: benchmark events recovered by MAD-LEO, MAD-LEO events present in the benchmark, and the reverse rate restricted to the benchmark's own time coverage; SWOT is omitted because the benchmark's records end in October 2022, before its launch. (b) Signed match-offset histogram; the cluster at $+1$~day reflects a day-of-year parsing difference in the benchmark's published dates. (c) TLE response-magnitude distributions of matched MAD-LEO events, unmatched events inside the benchmark's time coverage, and events outside it.}
\label{fig:external_validation}
\end{figure*}

\textbf{Annotation traceability.} The first group tests whether the labels are internally complete and externally corroborated. All 1,134 mission-reported annotations parse with source URLs, UTC timestamps, and explicit timing uncertainty; every record retains its verbatim source line, and day-of-year conversion is validated against the calendar of each year, including leap years. The external test matches the annotations against the independently compiled maneuver benchmark of Shorten et al. \cite{shorten2023wide} under a one-day tolerance (Figure~\ref{fig:external_validation}). In the forward direction, 687 of the 742 benchmark events (92.6\%) have a counterpart in MAD-LEO (panel a). The two label sets were compiled independently from different source material, so agreement at this level indicates that both describe the same underlying events. The matched-pair time offsets cluster at exactly $+1$~day (panel b; 582 of the 687 pairs fall within $\pm$1~h of $+1$~day). We verified this offset against the raw IDS/DORIS records: it is a day-of-year parsing difference in the benchmark's published dates and does not affect the released labels. In the reverse direction, 687 of the 1,134 mission-reported events appear in the benchmark. The benchmark ends in October 2022, so 392 events, among them every SWOT event, fall outside its span; restricted to the benchmark's own coverage, the reverse rate is 92.6\% (687 of 742 in-span events, a denominator that coincides in count with the number of benchmark events). The 55 in-span unmatched events concentrate where the benchmark is expected to be least complete: 19 have responses below 5~m at the catalog noise floor and 27 are tier-C windows at archive boundaries, with seven events in both groups. Panel (c) shows the same pattern in distribution: the unmatched in-span and out-of-span groups both carry heavier responses than the matched group (medians 33.4~m and 29.4~m against 13.2~m), so the unmatched remainder reflects benchmark recall at the noise floor and the relocation tail, not spurious annotations.

\textbf{Cross-source response consistency.} The second group tests whether the released response metric measures maneuver magnitude. Around each reported event, the TLE response is the change of the mean-element semi-major axis between the catalog epochs bracketing the analysis window, computed with Equation~\eqref{eq:tle-response}. With $n$ the catalog mean motion and $\mu$ the gravitational parameter,
\begin{equation}
\Delta a \;=\; a\!\left(n_{\mathrm{after}}\right) - a\!\left(n_{\mathrm{before}}\right), \qquad a(n) = \left(\frac{\mu}{n^{2}}\right)^{1/3},
\label{eq:tle-response}
\end{equation}
and angular element differences are unwrapped before evaluation so that no spurious $2\pi$ jumps enter. The precise-orbit response is brought to the same mean-element scale by the period-averaged conversion of Algorithm~\ref{alg:period-mean}, which suppresses the short-period, dominantly J$_2$, oscillation following classical mean-element theory \cite{brouwer1959solution,kozai1959motion}. The conversion is a first-order approximation to a rigorous mean-element transformation; the Kozai convention check below bounds the resulting scale error. Because the TLE response differences mean motion, itself a Kozai mean element, no short-period contamination enters the direct difference, and its noise floor is catalog fitting noise, quantified below. Forecast-error formulations, in which a catalog state is propagated to the next catalog epoch and compared against it, are a complementary detection-oriented metric and are not required for the magnitude estimates released here.

\begin{figure*}[!t]
\centering
\includegraphics[width=\textwidth]{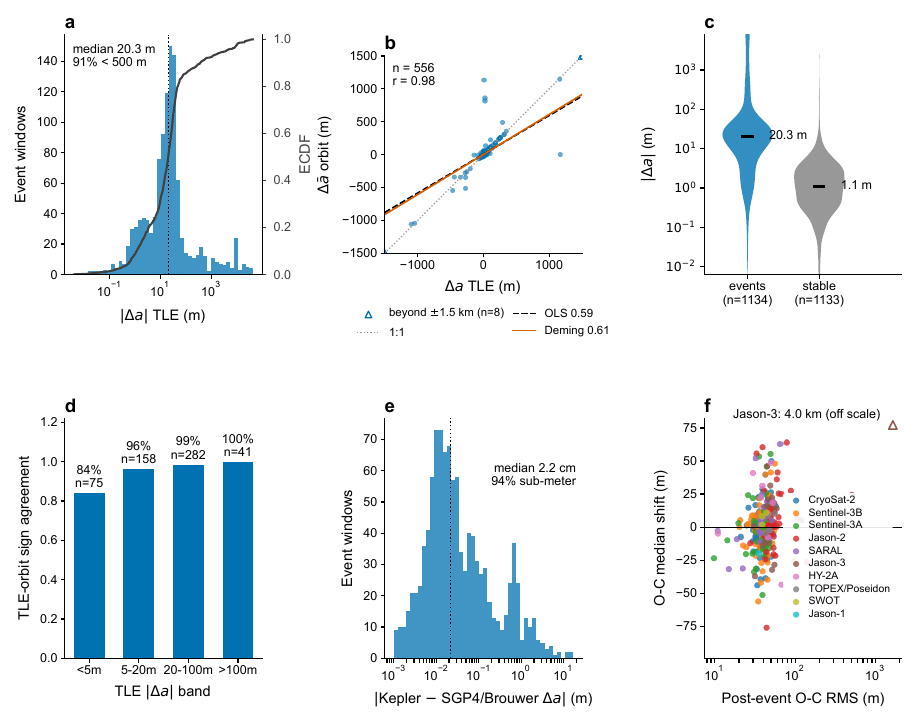}
\caption{Event response of the mission-reported subset. (a) Distribution and empirical cumulative distribution of the TLE mean-element response across all 1,134 event windows. (b) Same-scale comparison of TLE and period-averaged precise-orbit responses for the 556 dual-computable windows, with OLS and Deming fits. (c) Specificity against the stable-window negative controls. (d) Sign agreement between TLE and orbit estimates by response magnitude. (e) Kozai convention comparison: histogram of the per-event difference between the Kepler-path and SGP4 Brouwer-path semi-major-axis responses (median 2.2~cm, 94\% sub-meter). (f) SLR geometric observed-minus-computed residual shift versus post-event RMS for the computable event windows.}
\label{fig:event_response}
\end{figure*}

A TLE response is computable for all 1,134 windows, with a median $|\Delta a|$ of 20.3~m and 91\% of events below 500~m. Figure~\ref{fig:event_response} collects the checks on this estimator. Panel (a) shows the response distribution, a station-keeping population centered on tens of meters with a relocation-class tail. Panel (b) gives the central consistency result: the two independent magnitude estimates on the same scale for the 556 dual-computable windows. Panels (c) and (d) test specificity. The no-event controls respond at a median of 1.1~m against the event median of 20.3~m (panel c), so the metric separates reported events from quiet periods by more than an order of magnitude. Cross-source sign agreement rises from 84\% below 5~m to 100\% above 100~m (panel d), as expected if disagreement between the two sources is confined to the catalog noise floor \cite{vallado2013improved}. Panels (e) and (f) carry the convention and laser-ranging checks described below.

For the 556 dual-computable windows, the two estimates agree with a Pearson correlation of 0.98 (Spearman rank correlation 0.89; Pearson correlation of the log magnitudes 0.76), with medians of 23.8~m (TLE) and 26.5~m (orbit) and an OLS slope of 0.59 (95\% bootstrap CI 0.50--0.67; Deming slope 0.61, CI 0.52--0.73, with the pre-registered noise ratio $\lambda = (1.1/24.0)^2$, the squared ratio of the median response noise floors of the two estimators). Two checks constrain the origin of the sub-unity slope. First, a synthetic injection experiment on a quiet Sentinel-3A precise-orbit arc (released as \texttt{attenuation\_demo.csv}) shows that both estimators recover a strictly impulsive semi-major-axis step exactly, and that the plausible dilution mechanisms, period averaging and catalog bracket spacing acting on a finite-duration (40~h) ramp, would attenuate the TLE-scale estimator more strongly than the orbit estimator (implied orbit-to-TLE ratio 1.46 at the observed median bracket spacing of 46.8~h); dilution would therefore push the slope above unity, the opposite of what is observed. Second, magnitude-banded refits (released in \texttt{slope\_estimation\_comparison.csv}) show that the pooled slope is variance-weighted by the large-maneuver tail: the OLS slope is 1.06 in the 100--1,000~m band (median orbit-to-TLE ratio 1.01) but 0.61 above 1~km (median ratio 0.79, $n=12$), where the 24~m TLE noise floor is three orders of magnitude below the signal and errors-in-variables attenuation is negligible (for the pooled fit, the attenuation-corrected slope 0.58992 differs from the OLS slope 0.58985 by 0.01\%). The sub-unity pooled slope therefore reflects a systematic scale difference between the two estimators at the largest maneuvers, not noise attenuation. Per-satellite slopes scatter between 0.43 and 1.15 around the pooled value, and the median per-event orbit-to-TLE ratio above 10~m is 1.02, so the two estimators agree in distribution across the magnitude range even where the pooled regression slope departs from unity.

Three further checks bound the remaining systematic effects. The bracket spacing contributes negligibly through dynamics: in the released TLE histories, the median two-day drag decay of the semi-major axis is about 2~m for the lowest-altitude target (CryoSat-2, median altitude $\approx$725~km) and falls to centimeters at the 1,300~km altitudes of the Jason-family orbits, in every case well below the 24~m response noise floor. Within the $|\Delta a| < 20$~m half of the sample, per-window agreement degrades (Pearson 0.08, Spearman 0.62), so the two estimators should be read as consistent in distribution, not window by window, near the catalog noise floor. Recomputing every response under a symmetric $\pm$24~h window instead of the release $-6/+24$~h definition changes the median $|\Delta a|$ by 0.3~m (1.5\%), with a median per-window difference of 0.43~m and 97.6\% sign agreement. A separate recomputation from the SGP4 Brouwer-path mean element reproduces the TLE semi-major-axis response to a median difference of 2.2~cm (94\% sub-meter; panel e), so the Keplerian conversion is convention-robust for the bulk of the release. A tail of 5.6\% of events exceeds 1~m (maximum 16~m, SARAL), attributable to the drag-term and eccentricity dependence of the two mean-element conventions; for those events the convention difference is comparable to the smallest event responses.

The stable-window controls bound the false-positive behavior of the metric. Of the 1,139 stable windows, 1,133 are computable (the six exceptions sit at the first grid point of each satellite's TLE series, where no bracketing catalog epoch exists), with a 90th-percentile response of 5.3~m, a 95th of 8.0~m, and a 99th of 26~m (panel c). The coverage-matched subset gives a median of 1.96~m (273 of 274 computable). Fourteen stable windows exceed 20~m (maximum 2.55~km); seven are SWOT windows from the early-2023 commissioning orbit-raise period, which the IDS histories do not cover. These windows are retained and flagged \texttt{suspect\_unreported\_maneuver}, so the response signal is specific to reported events and does not follow from evidence availability. Out-of-plane responses are negligible throughout the release: the median inclination change is 1.1~arcsec and the largest $0.11^\circ$, so no predominantly out-of-plane maneuver is present and the semi-major-axis response captures the maneuver population. The 45 events anchored at the reported operation end time show a much smaller median response (3.3~m against 20.7~m for first-impulse events), as expected for interval-level epochs dominated by early TOPEX/Poseidon records.

\begin{figure*}[!t]
\centering
\includegraphics[width=\textwidth]{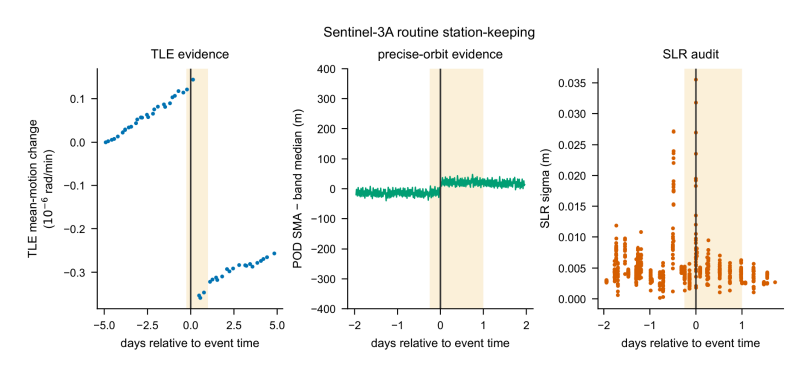}
\caption{Example event anatomy for Sentinel-3A, showing the coincident discontinuity in the TLE, precise-orbit, and SLR records around the mission-reported maneuver.}
\label{fig:anatomy_sentinel}
\end{figure*}

\begin{figure*}[!t]
\centering
\includegraphics[width=\textwidth]{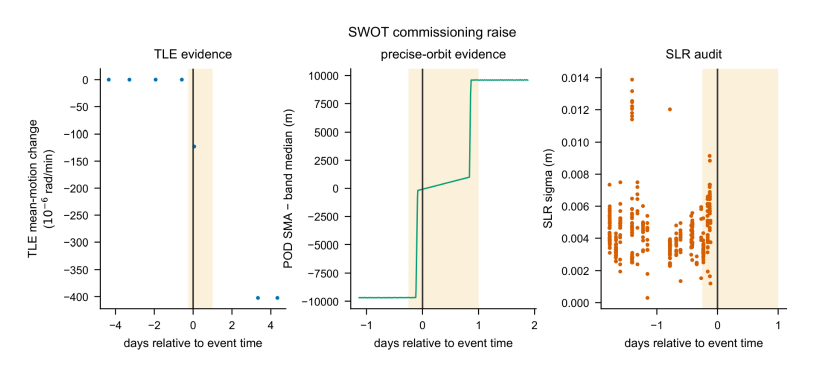}
\caption{Example event anatomy for SWOT, showing the coincident discontinuity in the TLE, precise-orbit, and SLR records around the mission-reported maneuver.}
\label{fig:anatomy_swot}
\end{figure*}

The two event anatomies (Figures~\ref{fig:anatomy_sentinel} and \ref{fig:anatomy_swot}) illustrate these checks at the level of a single event. At the reported epoch, the TLE mean element, the precise-orbit state series, and the independent laser ranges change together in sign and in time. The three processing chains share no step, so a coincident discontinuity in all three is direct evidence that the reported maneuver is present in the physical orbit. Sentinel-6A, whose orbit evidence is tracking coverage only, contributes no orbit-derived quantities and enters none of the dual-computable, orbit-response, state-estimation, or fusion analyses.

\begin{algorithm}[t]
\caption{Osculating orbit series $\Rightarrow$ period-averaged event response}
\label{alg:period-mean}
\begin{algorithmic}[1]
  \Require Osculating semi-major-axis series $\{a_k\}$ in the twelve-hour bands bracketing the analysis window; sampling step $\Delta t$
  \Ensure Period-averaged response $\Delta\bar a$ on the mean-element scale
  \Statex \textbf{Step 1 (unwrap).} Unwrap all angular element series to remove $2\pi$ discontinuities before any averaging.
  \Statex \textbf{Step 2 (nodal period).} Estimate the nodal period from the Keplerian relation
  \begin{equation*}
  T = 2\pi\sqrt{\bar a^{3}/\mu},
  \end{equation*}
  with $\bar a$ the band-averaged semi-major axis (the Keplerian period approximates the nodal period to better than 1\% for these near-circular J$_2$-dominated orbits).
  \Statex \textbf{Step 3 (period averaging).} Apply a centered rolling mean of width $T$ to the osculating series,
  \begin{equation*}
  \tilde a_k = \frac{1}{2m+1}\sum_{j=-m}^{m} a_{k+j}, \qquad m = \left\lfloor \frac{T}{2\,\Delta t} \right\rfloor ,
  \end{equation*}
  so that short-period oscillations cancel over complete cycles; at the band edges the average uses the available half window rather than truncating.
  \Statex \textbf{Step 4 (band medians and difference).} Take the median of $\tilde a_k$ within each bracketing band and report $\Delta\bar a = \mathrm{median}_{\mathrm{after}}(\tilde a) - \mathrm{median}_{\mathrm{before}}(\tilde a)$; a window is computable only if both bands hold at least two state samples (at least one period-averaged value each).
\end{algorithmic}
\end{algorithm}

\textbf{SLR as sparse audit.} The third group tests the laser-ranging evidence itself. SLR normal points bracket 542 event windows densely enough for before-and-after statistics, and the median shift in normal-point precision across these windows is 2.6~mm against a typical precision of 4--14~mm \cite{pearlman2002international} (Figure~\ref{fig:dataset}e): no systematic precision change accompanies reported maneuvers. A geometric orbit-referenced audit then compares each observed range against the range computed from the shipped precise-orbit states. Across the 296 windows with at least three observations on each side, the median observed-minus-computed RMS is 41~m, dominated by systematic terms deliberately left uncorrected in this audit (elevation-dependent tropospheric delay, which can reach the ten-meter level at low elevations, and center-of-mass offsets). The median before-to-after shift of the windowed median residual is 0.23~m (Figure~\ref{fig:event_response}f), so the precise-orbit products are corroborated by independent laser-ranging observations around reported events. Four windows exceed 1~km RMS and are disclosed: a Jason-1 window whose residuals are systematically offset by $-32$~km over ten observations (a pattern pointing to a time-tagging or epoch-convention problem in the source normal points; the median-based statistics are unaffected), a Jason-3 window at 4~km corresponding to a documented multi-day orbit degradation, and CryoSat-2 and TOPEX windows at 1.1 and 1.0~km at archive-boundary epochs. Every window lacking sufficient SLR coverage carries a machine-readable missing-source identifier in the released alignment table.

\begin{figure*}[!t]
\centering
\includegraphics[width=\textwidth]{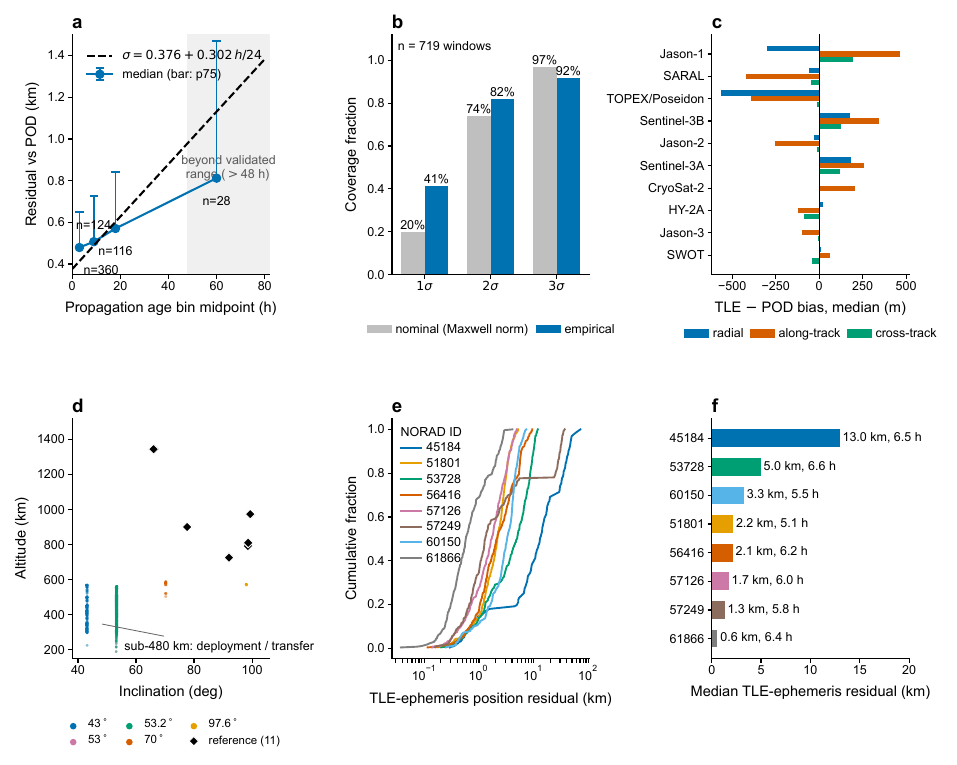}
\caption{Cross-source consistency audits. (a) SGP4 propagation residual curve against precise-orbit states, with the uncertainty model of Equation~\eqref{eq:sigma-model} overlaid. (b) Sigma-model calibration: Maxwell-norm nominal versus empirical coverage at 1$\sigma$, 2$\sigma$, and 3$\sigma$ (the bounded quantity is a three-dimensional norm, so the Maxwell reference, not the one-dimensional Gaussian, is the apples-to-apples nominal). (c) Systematic TLE-minus-POD biases per target in radial, along-track, and cross-track components. (d) Inclination--altitude distribution of the cataloged Starlink satellites in the released TLE snapshot, colored by plane family, with the eleven reference missions overlaid at their median orbital regimes; the sub-480~km population reflects deployment and transfer orbits. (e) Per-satellite empirical cumulative distributions of the Starlink TLE--ephemeris position residual of Equation~\eqref{eq:tle-ephemeris-residual}. (f) Per-satellite median Starlink TLE--ephemeris residual with median propagation age annotated.}
\label{fig:consistency}
\end{figure*}

\textbf{Operational subset boundary.} The fourth group tests the Starlink slice and the claims it can support. Temporal-overlap and schema-consistency audits document the intervals jointly covered by the ephemerides and the TLE records. The released TLE snapshot reproduces the constellation's shell architecture (Figure~\ref{fig:consistency}d), with inclination bands at 43.0$^\circ$ (2,125 satellites, 26 RAAN planes), 53.0$^\circ$ (1,311; 53), 53.2$^\circ$ (2,619; 22), 70$^\circ$ (397; 23), and 97.6$^\circ$ (232; 5), and with the altitude spread of the 43$^\circ$ shell reflecting ongoing deployment. The ephemeris slice keeps its nominal sixty-second cadence for every shell.

As a strict cross-source check, the nearest prior catalog TLE is propagated with SGP4 and converted from TEME to GCRS before comparison with the MEME ephemeris (Mean Equator Mean Equinox of the J2000.0 epoch, following the operator's published ephemeris format),
\begin{equation}
r(t) \;=\; \bigl\lVert\, \mathcal{T}_{\mathrm{TEME}\rightarrow\mathrm{GCRS}}\!\left(\mathbf{x}_{\mathrm{SGP4}}(t)\right) - \mathbf{x}_{\mathrm{eph}}(t) \,\bigr\rVert_{2},
\label{eq:tle-ephemeris-residual}
\end{equation}
evaluated at ten-minute spacing for a fixed-seed sample of eight satellites, restricted to states whose nearest prior catalog TLE is at most twelve hours old. This age cap covers 38--50\% of each sampled satellite's epochs and selects comparatively fresh catalog states, so the quoted medians are optimistic for the full slice. Per-satellite median residuals span 0.58 to 13.0~km at median propagation ages of five to seven hours (Figure~\ref{fig:consistency}e, f), of the same order as published Starlink SGP4 errors of 1--10~km; the published figures correspond to roughly one day of propagation, whereas the median ages here are five to seven hours, and the upper end of the observed range (13~km) belongs to the maneuver-contaminated satellite discussed below. A frame-sensitivity check supports the adopted convention: reading the ephemerides in an of-date frame shifts per-satellite median residuals to 30--40~km, while the J2000.0 reading reproduces the published processing chain to below one meter. Under this convention the frame transformations are exact, so the residuals are dominated by SGP4 propagation error at the corresponding TLE age and by unmodeled maneuvers inside the slice. The upper tail (99th percentile 46~km, maximum 71~km) is concentrated in two satellites over contiguous time intervals and is not age-driven (Pearson residual--age correlation 0.19). For NORAD 45184 the released TLE records show an abrupt 4.3~km semi-major-axis decrease at 22:00~UTC on 26 November 2024, about fifteen times its median quiet-period drag decay of 0.28~km per 8~h at 392~km altitude: an actual maneuver crosses the propagation interval. The released TLE table itself indicates orbit-change activity inside the slice: 10,498 consecutive-epoch semi-major-axis steps exceeding 100~m across 2,909 satellites (median 338~m, 57\% from the sub-480~km deployment and transfer population, where drag alone contributes changes of comparable size). No validated maneuver labels are provided for this subset.

\textbf{Harmonized state estimation.} The fifth group tests the harmonization layer, which estimates the orbit state of each event epoch from every available source. Measured against precise-orbit states, SGP4 propagation residuals grow from 0.48~km below six hours of propagation age to 0.86~km at twenty-four to forty-eight hours with edge-extrapolated windows excluded, or 0.98~km when these windows are included (Figure~\ref{fig:consistency}a). Growth with propagation age is the expected behavior of a mean-element catalog fit propagated away from its epoch, and the measured curve quantifies it for these missions. Beyond 48~h the binned curve turns non-monotone on the few surviving windows and is not interpreted. The base level agrees with the LEO-class average catalog position uncertainty reported by Flohrer et al. \cite{flohrer2008assessment}, while the growth rate is specific to the eleven well-tracked geodetic and altimetry missions in this release. This curve defines the propagation-dependent uncertainty model
\begin{equation}
\sigma(\Delta t) \;=\; \sigma_{0} + g\,\frac{\Delta t}{24\,\mathrm{h}}, \qquad \sigma_{0} = 0.376~\mathrm{km},\;\; g = 0.302~\mathrm{km},
\label{eq:sigma-model}
\end{equation}
fitted on bin-median residuals so that rare extrapolation outliers do not dominate; the linear form is a local approximation valid to 48~h. The orbit interpolation error budget, measured by decimation and recovery, stays at the millimeter level for DORIS SP3 products and at the micrometer level for the Sentinel-3 and SWOT products, with the Jason-3 OGDR product at the sub-millimeter level, well below the maneuver signal.

Per-target systematic biases between TLE and precise-orbit states reach hundreds of meters and are released in radial, along-track, and cross-track components for user-side correction (Figure~\ref{fig:consistency}c). The median absolute components are 296~m along-track, 93~m cross-track, and 68~m radial, and their ordering matches the catalog error budget of Flohrer et al., in which along-track errors dominate. In the leave-POD-out evaluation, the model of Equation~\eqref{eq:sigma-model} is fitted on four satellites (Sentinel-3A/3B, Jason-3, SARAL) and evaluated on disjoint targets. Three-sigma coverage is 95.5\% on the fit group, 96.9\% on the contemporary evaluation target (CryoSat-2), and 84.2\% on the historical evaluation targets (Jason-1/2, TOPEX/Poseidon, HY-2A, SWOT, grouped because none contributed to the fit and their orbit products come from earlier or independent processing chains). Users applying the model outside the fit group, in particular to historical-era targets, should treat the three-sigma bound as approximate and re-calibrate against their own residuals where possible. Because the bounded quantity is a three-dimensional position norm, the appropriate nominal reference is the Maxwell distribution, with nominal coverage of 19.9\%, 73.9\%, and 97.1\% at one, two, and three sigma. Against this reference the empirical coverage is 41\% and 82\% at one and two sigma, so the bound is conservative where it is most used and slightly below the reference at three sigma (Figure~\ref{fig:consistency}b). The model is a consistency bound, not a calibrated confidence interval.

\begin{figure*}[!t]
\centering
\includegraphics[width=\textwidth]{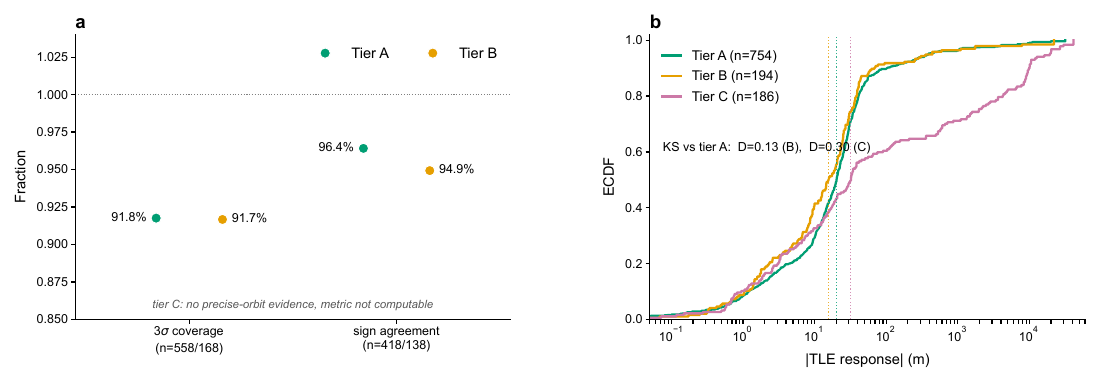}
\caption{Tier-stratified checks. (a) Orbit-referenced quality metrics by tier: three-sigma coverage and sign agreement are computable for tiers A and B only, because tier C windows carry no precise-orbit evidence. (b) Empirical cumulative distributions of the TLE response magnitude for all three tiers (dotted lines mark the tier medians); tier C carries the heaviest tail, with Kolmogorov--Smirnov distances against tier A of $D = 0.13$ (tier B) and $D = 0.30$ (tier C).}
\label{fig:tier_comparison}
\end{figure*}

\textbf{Tier stratification.} The sixth group tests whether the confidence tier of a window changes its measured quality. The two orbit-referenced metrics are computable only where precise-orbit evidence exists, so their comparison spans tiers A and B, the complete computable set (Figure~\ref{fig:tier_comparison}a). On this set the tiers are close: three-sigma coverage of 91.8\% (tier A) against 91.7\% (tier B) and sign agreement of 96.4\% against 94.9\%, with period-averaged orbit response medians of 22.0~m against 28.0~m. The TLE response is computable for every window and admits all three tiers (Figure~\ref{fig:tier_comparison}b). Its median rises from 20.2~m (tier A) and 15.5~m (tier B) to 31.8~m (tier C), and the tier-C distribution carries a heavy tail (75th percentile 1.87~km; Kolmogorov--Smirnov distance against tier A $D = 0.30$, $p = 3.5\times10^{-12}$), consistent with large maneuvers concentrating in the commissioning and disposal periods, where coverage is sparse. The tier A/B response distributions differ more subtly ($D = 0.13$, $p = 0.014$ for the TLE response; $D = 0.17$, $p = 0.004$ for the orbit response), consistent with the era composition of tier B. The tiers therefore separate coverage eras, not data quality: where a metric is computable at all, tier-B windows behave like tier-A windows. Every one of the 380 windows without full three-source coverage carries a machine-readable missing-source marker in the released alignment table.

\textbf{Evidence distribution completeness.} The seventh group tests the released evidence snapshots for sampling bias. Per-target TLE element distributions cluster tightly at the mission orbits, with inclination inter-quartile ranges below 0.033$^\circ$ for every target and altitude inter-quartile ranges of at most about 2~km outside documented orbit relocations. Epoch coverage spans 1992 to 2026 collectively across the eleven targets, each covering its complete mission period. Because tier composition varies with mission era, the tier-comparison checks are performed within each target; pooling across targets would confound orbital regime with tier. For each element $x$ and target $s$, the check reports the Kolmogorov--Smirnov statistic
\begin{equation}
D_{s} \;=\; \sup_{x} \left| F_{\mathrm{A},s}(x) - F_{\mathrm{B},s}(x) \right|,
\label{eq:tier-ks}
\end{equation}
evaluated on the per-event inclination, eccentricity, and semi-major-axis altitude at the nearest catalog epoch. Because a difference test cannot establish equivalence, formal tier equivalence is assessed separately with two one-sided tests (TOST) against tolerances pre-registered from engineering relevance (0.01$^\circ$ inclination, $7\times10^{-6}$ eccentricity, 0.05~km altitude), with Benjamini--Hochberg control across all element--target--pair comparisons. The verdicts confirm inclination equivalence in 14 of 19 within-target comparisons, and within-target inclination shifts between tiers are at most 0.033$^\circ$ (largest for HY-2A, whose tier-B epochs coincide with documented orbit maintenance). The eccentricity and altitude tolerances are exceeded in 24 of the 38 within-target comparisons, and every exceedance coincides with a documented mission phase, the largest being the 2018 relocation of Jason-2 (26.8 km), the 1992 commissioning of TOPEX (7.1 km), and the tier-era split of HY-2A (2.1 km). These shifts amount to at most 0.35\% of the semi-major axis and reflect mission history rather than evidence quality. On the inclination dimension the tier distributions are statistically equivalent.

The quality-control process itself identified three data issues that are documented in the release: the mixed time scales of the DORIS SP3 products, the legacy pre-CRD laser ranging format used by the earlier missions, and the boundary effects of SP3 arc coverage.

\section*{Usage Notes}

\textbf{Choosing a subset and an evidence level.} The mission-reported subset carries the validated maneuver labels and is the intended basis for maneuver-related training and evaluation. Its three evidence sources play different roles. TLE records are complete for every event and support the maneuver-magnitude analysis. Precise orbit products cover the windows of tiers A and B and provide the independent magnitude estimate. SLR normal points are sparse and serve as an event-signature audit, not as an orbit-accuracy reference. The operational subset contains operator-published predicted ephemerides and no validated maneuver labels. It is intended for operational-transfer and domain-shift studies in which methods developed on the reference labels are examined at constellation scale, and it must not be used as maneuver ground truth.

\textbf{Evaluation design.} Confidence tiers record evidence completeness, so analyses that require three-source corroboration should restrict themselves to tier A, tier B remains quantitatively reliable on its available sources, and tier C is intended for conservative use or manual review. The release provides no recommended train, validation, or test partition. Users who need partitions can derive them from the per-satellite and per-window metadata. Partitioning by satellite avoids leakage from satellite-specific signatures, while partitioning by time tests extrapolation within a mission, and the chosen construction should be stated. For evaluation we recommend the coverage-matched stable subset (274 windows), whose evidence coverage follows that of the events. The full stable set (1,139 windows) deliberately spans complete mission histories and should be used when temporal coverage matters more than coverage balance.

\textbf{Label-domain boundary.} The labeled events come exclusively from eleven DORIS-equipped geodetic and altimetry missions, whose maneuvers are dominated by in-plane, centimetre-per-second-class station keeping (median semi-major-axis response 20.3~m, equivalent to roughly 1~cm/s for a single-impulse along-track burn, $\Delta v = (n/2)\,\Delta a$ evaluated per event with each event's own mean motion). Transfer of methods trained on this label set to large impulsive or avoidance-class maneuvers has not been validated in this release. Because the response measures are based on the semi-major axis, they are insensitive to pure out-of-plane (inclination) maneuvers. The released events show no predominantly out-of-plane component (median inclination change 1.1~arcsec, maximum $0.11^\circ$), consistent with the mission class. For Sentinel-6A, whose orbit evidence is daily GNSS tracking coverage rather than a processed orbit product, no orbit-derived response exists. Include its windows in label-level analyses but exclude them from any metric that requires precise-orbit states.

\textbf{Reporting conventions.} When reporting results obtained with MAD-LEO, state the subset and evidence sources used, the confidence-tier selection (strict for tier A, relaxed for tiers A and B, audit for all), and the partition construction. Two magnitude estimators are provided per window, the TLE mean-element change and the period-averaged orbit change. They agree at a correlation of 0.98 on the 556 dual-computable windows but differ systematically at the large-maneuver end, so the estimator used should be stated. If several configurations are compared, results should be reported for each configuration separately rather than pooled. All timestamps are ISO-8601 UTC, positions are given in meters, velocities in meters per second, and angles in radians. Preprocessing applied to the released tables, such as filtering or aggregation, should be documented so that results remain reproducible against the release.

\textbf{Reading the caveats correctly.} Five conventions must not be misread. The released precise-orbit velocities are Earth-fixed (rotating-frame) values, and the $\omega \times r$ term must be added before any inertial-frame use. The Sentinel-6A orbit evidence is a daily GNSS tracking product rather than a finished orbit. The SLR precision fields quantify the spread of the measurements within a normal point and are not orbit residuals. The harmonization sigma model is a consistency bound and not a calibrated confidence interval. Candidate rows that appear in audit tables are evidence artifacts, not labels, and the public label vocabulary is limited to \texttt{event}, \texttt{no\_event}, and \texttt{ignore}. Rows flagged \texttt{qc\_status} or \texttt{below\_surface} in the evidence tables are retained verbatim for transparency and should be filtered before computing statistics. Windows with sparse evidence coverage retain machine-readable gap reasons, which should be consulted before missing coverage is interpreted as an event property.

\textbf{Reproduction.} Reproduction starts from the archived raw inputs or from the released snapshots. The commands for regenerating the normalized sources, the alignment audit, the response tables, and the figure set are documented in the code repository.

\section*{Code availability}

The dataset was constructed using open-source software. The complete pipeline is available at \url{https://github.com/sjtugzx/madLEO}.

\section*{Acknowledgements}
This work was supported by the Natural Science Foundation of Shanghai (Grant No. 25ZR1402471) and the Shanghai Sailing Program (Grant No. 24YF2743300).

\bibliography{sample}

\section*{Author contributions statement}
Zhixin Guo led conceptualization, methodology design, software development, data validation, data analysis, and writing (original draft), and contributed to writing (review \& editing). Qi Shi contributed to conceptualization, data curation, investigation, and writing (review \& editing). Xiaofan Xu led project administration and supervision, and contributed to conceptualization, funding acquisition, and writing (review \& editing). Linqiang Ge contributed to project administration, supervision, and funding acquisition. Hua Zhu contributed to project administration, supervision, and funding acquisition. Bendian Nie and Liyan Ben contributed to resources, data curation, and writing (review \& editing). Yuanrui Zhao contributed to visualization, data curation, and writing (review \& editing). Xiaohan Li contributed to investigation, validation, and writing (review \& editing). All authors reviewed and approved the final manuscript.

\section*{Competing interests}
The authors declare no competing interests.

\end{document}